\documentclass[cits]{PoS}
\usepackage{epsfig}
\usepackage{amsmath}

\title{Perturbative stability of the QCD predictions for the ratio $R=F_L/F_T$ and azimuthal asymmetry in heavy-quark leptoproduction}

\ShortTitle{Perturbative stability of the QCD predictions for $R=F_L/F_T$ and azimuthal asymmetry}

\author{\speaker{N. Ya. Ivanov}\\%
        Yerevan Physics Institute, Alikhanian Brs. 2, Yerevan 0036, Armenia\\
        E-mail: \email{nikiv@yerphi.am}}

\abstract{
\noindent We analyze the perturbative and parametric stability of the QCD predictions 
for the Callan-Gross ratio $R(x,Q^2)=F_L/F_T$  and azimuthal $\cos(2\varphi)$ asymmetry 
in heavy-quark leptoproduction.
Our analysis shows that large radiative corrections to the structure functions cancel 
each other in their ratio $R(x,Q^2)$ and azimuthal asymmetry  with good accuracy.
As a result, the NLO contributions to the Callan-Gross ratio  and $\cos(2\varphi)$ asymmetry 
are less than $10\%$ in a wide region of the variables $x$ and $Q^2$.
We provide compact analytic predictions for $R(x,Q^2)$ and asymmetry in the case of 
low $x\ll 1$. 
Simple formulae connecting the high-energy behavior of the Callan-Gross ratio and azimuthal asymmetry with the low-$x$ asymptotics of the gluon density in the target are derived. 
It is shown that the obtained hadron-level predictions for $R(x,Q^2)$ and azimuthal asymmetry are stable at $x\ll 1$ under the DGLAP evolution of the gluon distribution function. 

Concerning the experimental aspects, we propose to exploit the observed perturbative
stability of the Callan-Gross ratio and $\cos(2\varphi)$ asymmetry in the extraction of the   structure functions from the corresponding reduced cross sections. 
In particular, our obtained analytic expressions simplify essentially the determination of $F_2^c(x,Q^2)$ and $F_2^b(x,Q^2)$  from available data of the H1 Collaboration. 
Our results will also be useful in extraction of the azimuthal asymmetries from the incoming 
and future data on heavy-quark leptoproduction.}

\FullConference{XXI International Baldin Seminar on High Energy Physics Problems \\
		 September 10-15, 2012\\
		 JINR, Dubna, Russia}

\begin{document}

\section{Introduction}
In the framework of perturbative quantum chromodynamics (QCD), the basic spin-averaged
characteristics of heavy-flavor photo- \cite{Ellis-Nason,Smith-Neerven},  electro- \cite{LRSN}, and hadro-production \cite{Nason-D-E-1,Nason-D-E-2,BKNS} are known exactly up to the next-to-leading order (NLO).\footnote{Some recent results concerning the ongoing computations of the  next-to-next-to-leading order (NNLO) corrections to the heavy-flavor hadroproduction are presented in Refs.~\cite{Czakon-Mitov-1,Czakon-Mitov-2}} Although these explicit results are widely used at present
for a phenomenological description of available data (for a review, see Ref.~\cite{R-Vogt}), the key question remains open: How to test the applicability of QCD at fixed order to heavy-quark production?  The basic theoretical problem is that the NLO corrections are sizeable; they increase the leading-order (LO) predictions for both charm and bottom production cross sections by approximately a factor of two. Moreover, soft-gluon resummation of the threshold Sudakov logarithms indicates that higher-order contributions can also be substantial. (For reviews, see
Refs.~\cite{Laenen-Moch,kid2}.) On the other hand, perturbative instability leads
to a high sensitivity of the theoretical calculations to standard uncertainties in the
input QCD parameters. The total uncertainties associated with the unknown values of the heavy-quark mass, $m$, the factorization and renormalization scales, $\mu _{F}$ and $\mu _{R}$, the asymptotic scale parameter $\Lambda_{\mathrm{QCD}}$ and the parton distribution functions (PDFs) are so large that one can only estimate the order of magnitude of the pQCD predictions for charm production cross sections in the entire energy range from the fixed-target experiments
\cite{Mangano-N-R,Frixione-M-N-R} to the RHIC collider \cite{R-Vogt}.

Since these production cross sections are not perturbatively stable, it is of special interest
to study those observables that are well-defined in pQCD. Nontrivial examples of such  observables were proposed in Refs.~\cite{we1,we2,we3,we4,we5,we6,we7,we8,we9}, where the azimuthal $\cos(2\varphi)$ asymmetry and Callan-Gross ratio $R(x,Q^2)=F_L/F_T$ in heavy-quark  leptoproduction were analyzed.\footnote{Well-known examples include the shapes of differential cross sections of heavy flavor production, which are sufficiently stable under radiative
corrections.}$^{,}$\footnote{Note also the paper \cite{Almeida-S-V}, where the
perturbative stability of the QCD predictions for the charge asymmetry in top-quark
hadroproduction has been observed.} In particular, the NLO soft-gluon corrections to the basic mechanism, photon-gluon fusion (GF), were calculated. It was shown that, contrary to
the production cross sections, the azimuthal $\cos(2\varphi)$ asymmetry in heavy-flavor photo- and leptoproduction is quantitatively well defined in pQCD: the contribution of the dominant
GF mechanism to the asymmetry is stable, both parametrically and perturbatively. Therefore, measurements of this asymmetry should provide a clean test of pQCD.

The perturbative and parametric stability of the GF predictions for the Callan-Gross ratio 
$R(x,Q^2)=F_L/F_T$ in heavy-quark leptoproduction was considered in Refs.~\cite{we7,we8,we9}. 
It was shown that large radiative corrections to the structure functions $F_T(x,Q^2)$ and 
$F_L(x,Q^2)$ cancel each other in their ratio $R(x,Q^2)$ with good accuracy. As a result, 
the next-to-leading order (NLO) contributions of the dominant GF mechanism to the Callan-Gross 
ratio are less than $10\%$ in a wide region of the variables $x$ and $Q^2$.

\begin{figure}[t]
\begin{center}
\mbox{\epsfig{file=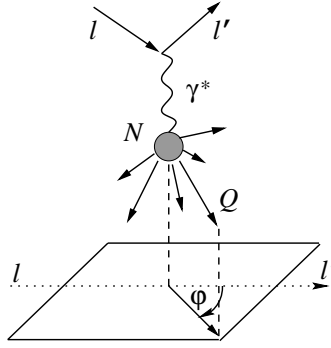,width=160pt}}
\caption{\label{Fg.1}\small Definition of the azimuthal angle $\varphi$ in the nucleon rest frame.}
\end{center}
\end{figure}

In the present paper, we continue the studies of perturbatively stable observables in heavy-quark leptoproduction,
\begin{equation}
\ell(l )+N(p)\rightarrow \ell(l -q)+Q(p_{Q})+X[\bar{Q}](p_{X}). \label{1}
\end{equation}
In the case of unpolarized initial states and neglecting the contribution of $Z$-boson exchange,
the azimuth-dependent cross section of the reaction (\ref{1}) can be written as
%\begin{eqnarray}
%\frac{\mathrm{d}^{3}\sigma_{lN}}{\mathrm{d}x\mathrm{d}Q^{2}\mathrm{d}\varphi }&=&\frac{2\alpha^{2}_{em}}{Q^4}
%\frac{y^2}{1-\varepsilon}\Bigl[ F_{T}( x,Q^{2})+ \varepsilon F_{L}(x,Q^{2}) \Bigr. \nonumber \\
%&+&\Bigl. \varepsilon F_{A}( x,Q^{2})\cos 2\varphi+
%2\sqrt{\varepsilon(1+\varepsilon)} F_{I}( x,Q^{2})\cos \varphi\Bigr], \label{2}
%\end{eqnarray}
\begin{equation}
\frac{\mathrm{d}^{3}\sigma_{lN}}{\mathrm{d}x\mathrm{d}Q^{2}\mathrm{d}\varphi }=\frac{2\alpha^{2}_{\mathrm{em}}}{Q^4}
\frac{y^2}{1-\varepsilon}\left[ F_{T}( x,Q^{2})+ \varepsilon F_{L}(x,Q^{2})
+ \varepsilon F_{A}( x,Q^{2})\cos 2\varphi+
2\sqrt{\varepsilon(1+\varepsilon)} F_{I}( x,Q^{2})\cos \varphi\right] \label{2}
\end{equation}
where $\alpha_{\mathrm{em}}$ is Sommerfeld's fine-structure constant, $F_{2}(x,Q^2)=2x(F_{T}+F_{L})$, the quantity $\varepsilon$ measures the degree of the longitudinal polarization of the virtual photon in the Breit frame \cite{dombey}, $\varepsilon=\frac{2(1-y)}{1+(1-y)^2}$, and the kinematic variables are defined by
\begin{eqnarray}
\bar{S}=\left( \ell +p\right) ^{2},\qquad &Q^{2}=-q^{2},\qquad &x=\frac{Q^{2}}
{2p\cdot q},  \nonumber \\
y=\frac{p\cdot q}{p\cdot \ell },\qquad \quad ~ &Q^{2}=xy\bar{S},\qquad &\xi=
\frac{Q^2}{m^2}.  \label{3}
\end{eqnarray}
In Eq.~(\ref{2}), $F_{T}\,(F_{L})$ is the usual $\gamma ^{*}N$ structure function describing
heavy-quark production by a transverse (longitudinal) virtual photon. The third structure function, $F_{A}$, comes about from interference between transverse states and is responsible for the $\cos2\varphi$ asymmetry which occurs in real photoproduction using linearly polarized photons. The fourth structure function, $F_{I}$, originates from interference between
longitudinal and transverse components \cite{dombey}. In the nucleon rest frame, the azimuth
$\varphi$ is the angle between the lepton scattering plane and the heavy quark production plane,
defined by the exchanged photon and the detected quark $Q$ (see Fig.~\ref{Fg.1}). The covariant
definition of $\varphi $ is
\begin{eqnarray}
\cos \varphi &=&\frac{r\cdot n}{\sqrt{-r^{2}}\sqrt{-n^{2}}},\qquad \qquad
\sin \varphi =\frac{Q^{2}\sqrt{1/x^{2}+4m_{N}^{2}/Q^{2}}}{2\sqrt{-r^{2}}%
\sqrt{-n^{2}}}~n\cdot \ell ,  \label{4} \\
r^{\mu } &=&\varepsilon ^{\mu \nu \alpha \beta }p_{\nu }q_{\alpha }\ell _{\beta },\qquad \ \
\qquad n^{\mu }=\varepsilon ^{\mu \nu \alpha \beta }q_{\nu }p_{\alpha }p_{Q\beta }.  \label{5}
\end{eqnarray}
In Eqs.~(\ref{3}) and (\ref{5}), $m$ and $m_{N}$ are the masses of the heavy quark and the target, respectively.

In this talk, we review the perturbative and parametric stability of the Callan-Gross ratio, $R(x,Q^{2})$, and azimuthal $\cos(2\varphi)$ asymmetry, $A(x,Q^{2})$, defined as
\begin{equation}\label{6}
R(x,Q^{2})=\frac{F_{L}}{F_{T}}(x,Q^{2}), \qquad \qquad A(x,Q^{2})=2x\frac{F_{A}}{F_{2}}(x,Q^{2}).
\end{equation}
First, we consider radiative corrections to the quantity $R(x,Q^2)$ using the explicit NLO results presented in \cite{LRSN,Blumlein}. Our calculations show that complete ${\cal O}(\alpha_{s}^2)$ corrections to $R(x,Q^2)$ (including both the photon-gluon, $\gamma ^{*}g\to Q\bar{Q}(g)$, and photon-(anti)quark, $\gamma ^{*}q\to Q\bar{Q}q$, fusion components) do not exceed 10$\%$ in the energy range $x>10^{-4}$. 

Then, we analyze the perturbative stability of the azimuthal $\cos(2\varphi)$ asymmetry, $A(x,Q^{2})$. Presently, the exact NLO predictions for the azimuth dependent structure function $F_{A}(x,Q^{2})$ are not available. For this reason, we use the so-called soft-gluon approximation to estimate the radiative corrections to $F_{A}(x,Q^{2})$. Our analysis shows that the NLO soft-gluon predictions for $A(x,Q^{2})$ affect the LO results by less than a few percent at $Q^2 \lesssim m^2$ and $x\gtrsim 10^{-2}$. 

In both cases, perturbative stability is mainly due to the cancellation of large radiative corrections to the structure functions $F_{L}$, $F_{T}$, $F_{A}$ and $F_{2}$ in their ratios, $R(x,Q^2)$ and $A(x,Q^{2})$, correspondingly. Note also that both the LO and NLO predictions for the Callan-Gross ratio and azimuthal asymmetry are sufficiently insensitive, to within ten percent, to standard uncertainties in the QCD input parameters $\mu_{F}$, $\mu_{R}$, $\Lambda_{\mathrm{QCD}}$ and PDFs. 

We conclude that, in contrast to the production cross sections, the ratios $R(x,Q^2)$ and $A(x,Q^{2})$ in heavy-quark leptoproduction are observables quantitatively well defined in pQCD. Measurements of these quantities in charm and bottom leptoproduction should provide a good test of the conventional parton model based on pQCD.

Since the ratios $R(x,Q^2)$ and $A(x,Q^{2})$ are perturbatively stable, it makes sense to provide the LO hadron-level predictions for these quantities in analytic form that may be useful in some applications. For this reason, we derive compact hadron-level LO predictions for the the Callan-Gross ratio and azimuthal asymmetry in the limit of low $x\to 0$. Assuming the low-$x$ asymptotic behavior of the gluon PDF to be of the type $g(x,Q^2)\propto 1/x^{1+\delta}$, we provide analytic result for the ratios $R(x\to 0,Q^2)$ and $A(x\to 0,Q^{2})$ for arbitrary values of the parameter $\delta$ in terms of the Gauss hypergeometric function.\footnote{The simplest case, $\delta =0$, has been studied in Ref.~\cite{kotikov}. The choice $\delta=1/2$ historically originates from the BFKL resummation of the leading powers of $\ln(1/x)$ \cite{BFKL1,BFKL2,BFKL3}.}  

In principle, the parameter $\delta$ is a function of $Q^2$ and this dependence is calculated 
using the DGLAP evolution equations \cite{DGLAP1,DGLAP2,DGLAP3}. However, our analysis shows that hadron-level predictions for $R(x\to 0,Q^2)$ and $A(x\to 0,Q^{2})$ are practically independent of $\delta$  in the entire region of $Q^2$ for $\delta > 0.2$. We see that the hadron-level predictions for $R(x\to 0,Q^2)$ and $A(x\to 0,Q^{2})$ are stable not only under the NLO corrections to the partonic cross sections, but also under the DGLAP evolution of the gluon PDF.

As to the experimental applications, we show that our compact LO formulae for $R(x\to 0,Q^2)$ conveniently reproduce the HERA results for $F_2^c(x,Q^2)$ and $F_2^b(x,Q^2)$ obtained by H1 Collaboration \cite{H1HERA1,H1HERA2} with the help of more cumbersome NLO estimations of $F_{L}(x,Q^2)$.
Our analytic predictions will also be useful in extraction of the azimuthal asymmetries from the incoming COMPASS results as well as from future data on heavy-quark leptoproduction at the proposed EIC \cite{EIC} and LHeC \cite{LHeC} colliders at BNL/JLab and CERN, correspondingly.

This paper is organized as follows. In Section~\ref{NLO}, we analyze the exact NLO results
for the Callan-Gross ratio. The soft-gluon contributions to $A(x,Q^{2})$ are investigated in Section~\ref{SGR}. The analytic LO results for the ratios $R(x,Q^{2})$ and $A(x,Q^{2})$ at low $x$ are discussed in Section~\ref{analytic}. 

\section{\label{NLO} Exact NLO predictions for the Callan-Gross ratio $R(x,Q^2)$}

\begin{figure}[b]
\begin{center}
\mbox{\epsfig{file=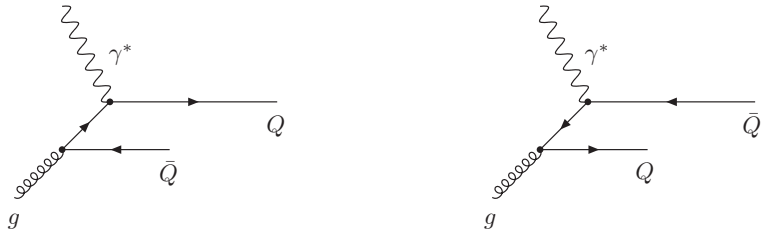,width=320pt}}
\end{center}
\caption{\label{Fg.2}\small Feynman diagrams of photon-gluon fusion at LO.}
\end{figure}

At leading order, ${\cal O}(\alpha_{\mathrm{em}}\alpha_{s})$, leptoproduction of heavy 
flavors proceeds through the photon-gluon fusion (GF) mechanism,
\begin{equation} \label{7}
\gamma ^{*}(q)+g(k_{g})\rightarrow Q(p_{Q})+\bar{Q}(p_{\bar{Q}}).
\end{equation}
The relevant Feynman diagrams are depicted in Fig.~\ref{Fg.2}.
The corresponding $\gamma ^{*}g$ cross sections, $\hat{\sigma}_{k,g}^{(0)}(z,\lambda)$ ($k=2,L,A,I$), have the form \cite{LW1}:
\begin{eqnarray}
\hat{\sigma}_{2,g}^{(0)}(z,\lambda)&=&\frac{\alpha_{s}}{2\pi}\hat{\sigma}_{B}(z)
\Bigl\{\left[(1-z)^{2}+z^{2}+4\lambda z(1-3z)-8\lambda^{2}z^{2}\right]
\ln\frac{1+\beta_{z}}{1-\beta_{z}}  \nonumber\\
&&-\left[1+4z(1-z)(\lambda-2)\right]\beta_{z}\Bigr\},  \nonumber\\
\hat{\sigma}_{L,g}^{(0)}(z,\lambda)&=&\frac{2\alpha_{s}}{\pi}\hat{\sigma}_{B}(z)z
\Bigl\{-2\lambda z\ln\frac{1+\beta_{z}}{1-\beta_{z}}+\left(1-z\right)\beta_{z}\Bigr\},  \label{8} \\
\hat{\sigma}_{A,g}^{(0)}(z,\lambda)&=&\frac{\alpha_{s}}{\pi}\hat{\sigma}_{B}(z)z
\Bigl\{2\lambda\left[1-2z(1+\lambda)\right]\ln\frac{1+\beta_{z}}{1-\beta_{z}}+
(1-2\lambda)(1-z)\beta_{z}\Bigr\}, \nonumber  \\
\hat{\sigma}_{I,g}^{(0)}(z,\lambda)&=&0,  \nonumber
\end{eqnarray}
with $\hat{\sigma}_{B}(z)=(2\pi)^2e_{Q}^{2}\alpha_{\mathrm{em}}\,z/Q^{2}$, 
where $e_{Q}$ is the electric charge of quark $Q$ in units of the positron charge and  
$\alpha_{s}\equiv\alpha_{s}(\mu_R^2)$ is the strong-coupling constant. 
In Eqs.~(\ref{8}), we use the following definition of partonic kinematic variables:
\begin{equation}\label{9}
z=\frac{Q^{2}}{2q\cdot k_{g}},\qquad\lambda =\frac{m^{2}}{Q^{2}}, \qquad
\beta_{z}=\sqrt{1-\frac{4\lambda z}{1-z}}.
\end{equation}
The hadron-level cross sections, $\sigma_{k,GF}(x,Q^2)$ ($k=2,L,A,I$), corresponding to the GF subprocess, have the form
\begin{equation}\label{10}
\sigma_{k,GF}(x,Q^2)=\int_{x(1+4\lambda)}^{1}\mathrm{d}z\,g(z,\mu_{F})
\hat{\sigma}_{k,g}\left(x/z,\lambda,\mu_{F}\right),
\end{equation}
where $g(z,\mu_{F})$ is the gluon PDF of the proton. 

The leptoproduction cross sections $\sigma_{k}(x,Q^2)$ are related to the structure functions $F_{k}(x,Q^2)$ as follows:
\begin{eqnarray}
F_{k}(x,Q^2) &=&\frac{Q^{2}}{8\pi^{2}\alpha_{\mathrm{em}}x}\sigma_{k}(x,Q^2)
\qquad (k=T,L,A,I),  \nonumber \\
F_{2}(x,Q^2) &=&\frac{Q^{2}}{4\pi^{2}\alpha_{\mathrm{em}}}\sigma_{2}(x,Q^2), \label{11}
\end{eqnarray}
where $\sigma_{2}(x,Q^2)=\sigma_{T}(x,Q^2)+\sigma_{L}(x,Q^2)$.

At NLO, ${\cal O}(\alpha_{\mathrm{em}}\alpha_{s}^2)$, the contributions of both the photon-gluon, $\gamma ^{*}g\to Q\bar{Q}(g)$, and photon-(anti)quark, $\gamma ^{*}q\to Q\bar{Q}q$, fusion components are usually presented in terms of the dimensionless coefficient functions $c_{k}^{(n,l)}(z,\lambda)$ as
\begin{equation}\label{12}
\hat{\sigma}_{k}(z,\lambda,\mu^{2})=\frac{e_{Q}^{2}\alpha_{\mathrm{em}}\alpha_{s}}{m^{2}}
\left\{ c_{k}^{(0,0)}(z,\lambda)+ 4\pi\alpha_{s}\left[
c_{k}^{(1,0)}(z,\lambda)+c_{k}^{(1,1)}(z,\lambda)\ln\frac{\mu^{2}}{m^{2}}
\right]+{\cal O}(\alpha_{s}^2)\right\},
\end{equation}
where we identify $\mu=\mu_{F}=\mu_{R}$.

The coefficients $c_{k,g}^{(1,1)}(z,\lambda)$ and $c_{k,q}^{(1,1)}(z,\lambda)$ ($k=T,L,A,I$) of the $\mu$-dependent logarithms can be evaluated explicitly using renormalization group arguments \cite{Ellis-Nason,LRSN}. The results of direct calculations of the coefficient functions $c_{k,g}^{(1,0)}(z,\lambda)$ and $c_{k,q}^{(1,0)}(z,\lambda)$ ($k=T,L$) are presented in Refs.~\cite{LRSN,Blumlein}. Using these NLO predictions, we analyze the $Q^2$ dependence of the ratio $R(x,Q^2)=F_L/F_T$ at fixed values of $x$. 

\begin{figure}[t]
\begin{center}
\begin{tabular}{cc}
\mbox{\epsfig{file=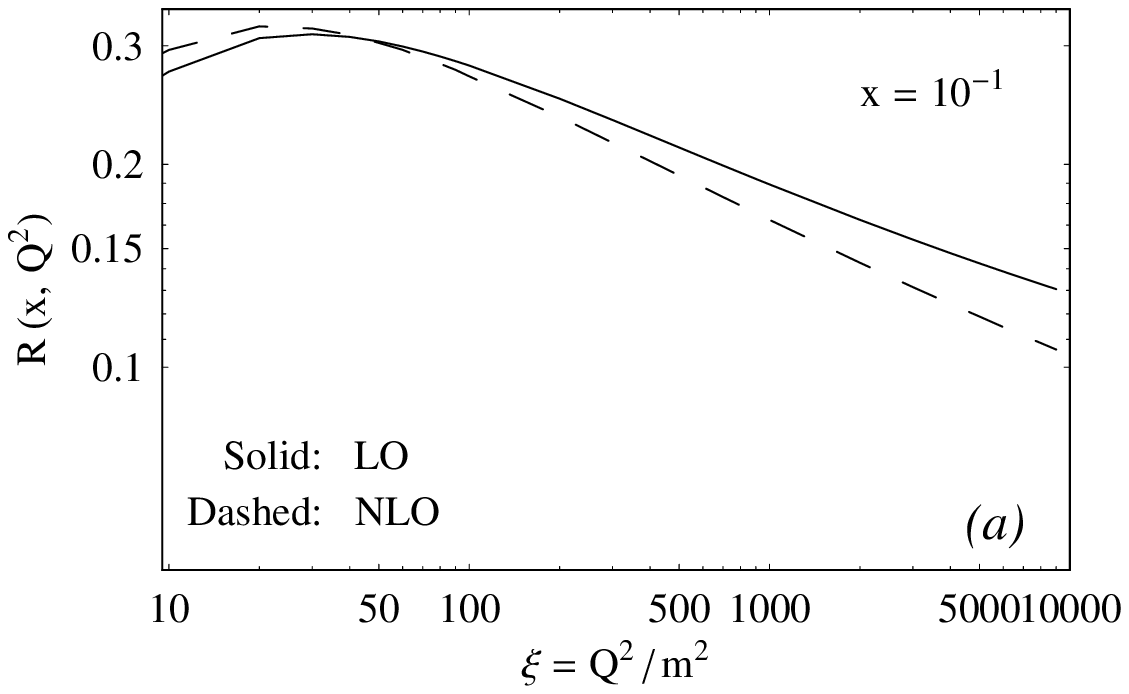,width=210pt}}
& \mbox{\epsfig{file=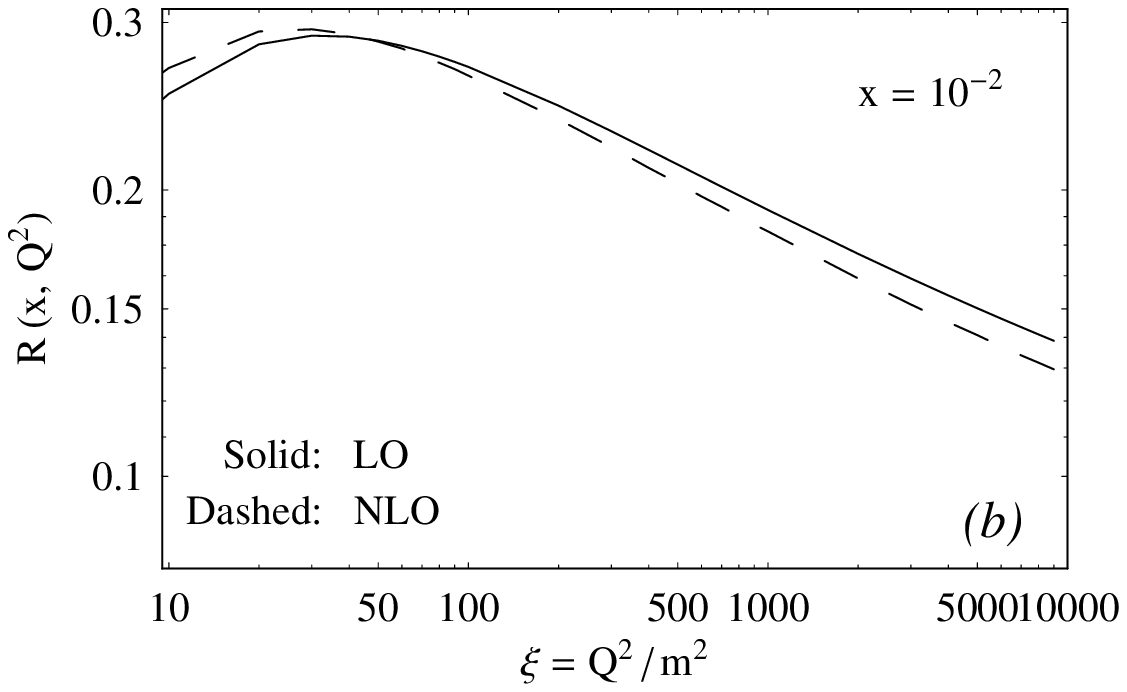,width=210pt}}\\
\mbox{\epsfig{file=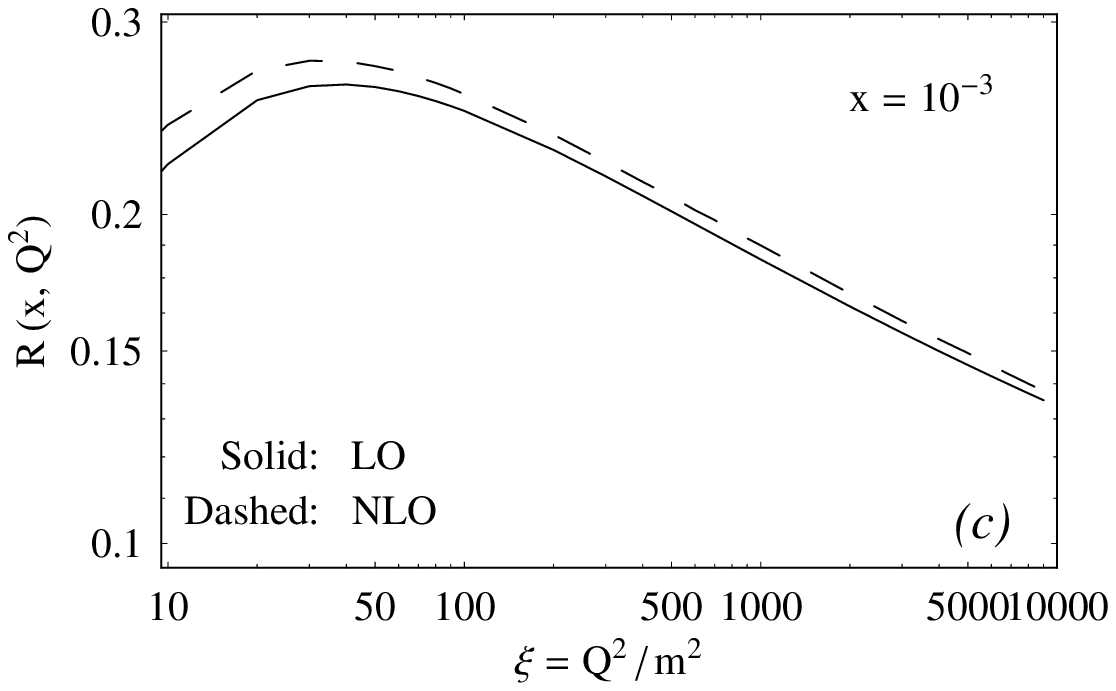,width=210pt}}
& \mbox{\epsfig{file=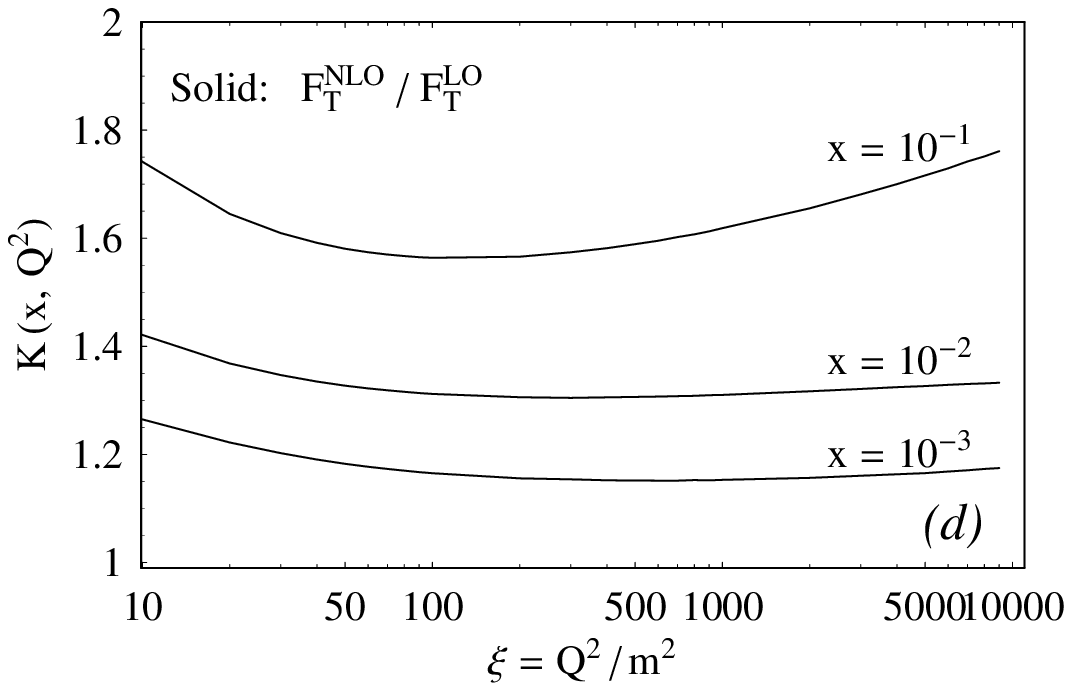,width=210pt}}\\
\end{tabular}
\caption{\label{Fg.3}\small $(a)$, $(b)$ and $(c)$ \emph{panels:} $Q^2$ dependence of the LO (solid curves) and NLO (dashed curves) predictions for the Callan-Gross ratio, $R(x,Q^2)=F_L/F_T$, in charm leptoproduction at $x=10^{-1}$, $10^{-2}$ and $10^{-3}$. \emph{$(d)$~panel:} $Q^2$ dependence of the $K$ factor for the transverse structure function,
$K(x,Q^2)=F_T^{\mathrm{NLO}}/F_T^{\mathrm{LO}}$, at the same values of $x$.}
\end{center}
\end{figure}

The panels $(a)$, $(b)$ and $(c)$ of Fig.~\ref{Fg.3} show the NLO predictions for Callan-Gross ratio $R(x,Q^2)$ in charm leptoproduction as a function of $\xi=Q^2 /m^2$ at $x=10^{-1}$, $10^{-2}$ and $10^{-3}$, correspondingly. In our calculations, we use the CTEQ6M parametrization of the PDFs together with the values $m_c=1.3$~GeV and $\varLambda=326$~MeV \cite{CTEQ6}.\footnote{Note that we convolve the NLO CTEQ6M distribution functions with both the LO and NLO partonic cross sections that makes it possible to estimate directly the degree of stability of the pQCD predictions under radiative corrections.} Unless otherwise stated, we use $\mu=\sqrt{4m_c^{2}+Q^{2}}$ throughout this paper.

For comparison, the panel $(d)$ of Fig.~\ref{Fg.3} shows the $Q^2$ dependence of the QCD
correction factor for the transverse structure function,
$K(x,Q^2)=F_T^{\mathrm{NLO}}/F_T^{\mathrm{LO}}$. One can see that sizable radiative corrections to the structure functions $F_T(x,Q^2)$ and $F_L(x,Q^2)$ cancel each other in their ratio $R(x,Q^2)=F_L/F_T$ with good accuracy. As a result, the NLO contributions to the ratio $R(x,Q^2)$ are less than $10\%$ for $x > 10^{-4}$.

Another remarkable property of the Callan-Gross ratio closely related to fast
perturbative convergence is its parametric stability.\footnote{Of course, parametric
stability of the fixed-order results does not imply a fast convergence of the
corresponding series. However, a fast convergent series must be parametrically stable. In
particular, it must exhibit feeble $\mu_{F}$ and $\mu_{R}$ dependences.}
Our analysis shows that the fixed-order predictions for the ratio $R(x,Q^2)$ are less sensitive to standard uncertainties in the QCD input parameters than the corresponding ones for the production cross sections. For instance, sufficiently above the production threshold, changes of
$\mu$ in the range $(1/2)\sqrt{4m_{c}^{2}+Q^{2}}<\mu <2 \sqrt{4m_{c}^{2}+Q^{2}}$
only lead to $10\%$ variations of $R(x,Q^{2})$ at NLO. For comparison, at
$x=0.1$ and $\xi = 4.4$, such changes of $\mu$ affect the NLO predictions for the
quantities $F_{T}(x,Q^2)$ and $R(x,Q^{2})$ in charm leptoproduction by more than $100\%$
and less than $10\%$, respectively.

Keeping the value of the variable $Q^{2}$ fixed, we analyze the dependence of the pQCD
predictions on the uncertainties in the heavy-quark mass. We observe that changes of the charm-quark mass in the interval 1.3~GeV${}<m_{c}<1.7$~GeV affect the Callan-Gross ratio by 2\%--3\% at $Q^{2}=10$~GeV$^2$ and $x<10^{-1}$. The corresponding variations of the structure functions $F_T(x,Q^2)$ and $F_L(x,Q^2)$ are about 20\%.
We also verify that the recent CTEQ versions \cite{CTEQ6,CT10} of the PDFs lead to
NLO predictions for $R(x,Q^{2})$ that coincide with each other with an accuracy of about
$5\%$ at $10^{-3}\leq x< 10^{-1}$.

\section{\label{SGR} Soft-gluon corrections to the azimuthal asymmetry $A(x,Q^{2})$ at NLO}

Presently, the exact NLO predictions for the azimuth dependent structure function $F_{A}(x,Q^{2})$ are not available. For this reason, we consider the NLO predictions for the azimuthal $\cos(2\varphi)$ asymmetry within the soft-gluon approximation. For the reader's convenience, we collect the final results for the parton-level GF cross sections to the next-to-leading logarithmic (NLL) accuracy. More details may be found in
Refs.~\cite{Laenen-Moch,we2,we4,we7}.

At NLO, photon-gluon fusion receives contributions from the virtual ${\cal
O}(\alpha_{\mathrm{em}}\alpha_{s}^{2})$ corrections to the Born process~(\ref{7}) and
from real-gluon emission,
\begin{equation}  \label{13}
\gamma ^{*}(q)+g(k_{g})\rightarrow Q(p_{Q})+\bar{Q}(p_{\bar{Q}})+g(p_{g}).
\end{equation}
The partonic invariants describing the single-particle inclusive (1PI) kinematics are
\begin{eqnarray}
s^{\prime }&=&2q\cdot k_{g}=s+Q^{2}=\zeta S^{\prime },\qquad \qquad \, t_{1}=\left(
k_{g}-p_{Q}\right) ^{2}-m^{2}=\zeta T_{1},  \nonumber \\
s_{4}&=&s^{\prime }+t_{1}+u_{1},\qquad \qquad \qquad \qquad ~~ u_{1}=\left(
q-p_{Q}\right) ^{2}-m^{2}=U_{1}, \label{14}
\end{eqnarray}
where $\zeta$ is defined through $\vec{k}_{g}= \zeta\vec{p}\,$ and $s_{4}$ measures the
inelasticity of the reaction (\ref{13}). The corresponding 1PI hadron-level variables
describing the reaction (\ref{1}) are
\begin{eqnarray}
S^{\prime }&=&2q\cdot p=S+Q^{2},\qquad \qquad T_{1}=\left( p-p_{Q}\right)
^{2}-m^{2},  \nonumber \\
S_{4}&=&S^{\prime }+T_{1}+U_{1},\qquad \qquad \quad U_{1}=\left( q-p_{Q}\right)
^{2}-m^{2}. \label{15}
\end{eqnarray}

The exact NLO calculations of unpolarized heavy-quark production \cite{Ellis-Nason,Smith-Neerven,LRSN,Nason-D-E-1} show that, near the partonic
threshold, a strong logarithmic enhancement of the cross sections takes place in the
collinear, $|\vec{p}_{g,T}|\to 0$, and soft, $|\vec{p}_{g}|\to 0$,
limits. This threshold (or soft-gluon) enhancement is of universal nature in perturbation
theory and originates from an incomplete cancellation of the soft and collinear
singularities between the loop and the bremsstrahlung contributions. Large leading and
next-to-leading threshold logarithms can be resummed to all orders of the perturbative
expansion using the appropriate evolution equations
\cite{Contopanagos-L-S}. The analytic results for the resummed
cross sections are ill-defined due to the Landau pole in the coupling constant
$\alpha_{s}$. However, if one considers the obtained expressions as generating functionals
and re-expands them at fixed order in $\alpha_{s}$, no
divergences associated with the Landau pole are encountered.

Soft-gluon resummation for the photon-gluon fusion was performed in
Ref.~\cite{Laenen-Moch} and confirmed in Refs.~\cite{we2,we4}. To NLL accuracy, the
perturbative expansion for the partonic cross sections,
$\mathrm{d}^{2}\hat{\sigma}_{k}(s^{\prime},t_{1},u_{1})/(\mathrm{d}t_{1}\,
\mathrm{d}u_{1})$ ($k=T,L,A,I$), can be written in factorized form as
\begin{equation}  \label{16}
s^{\prime 2}\frac{\text{d}^{2}\hat{\sigma}_{k}}{\text{d}t_{1}\text{d}u_{1}}( s^{\prime
},t_{1},u_{1}) =B_{k}^{\text{{\rm Born}}}( s^{\prime },t_{1},u_{1})\left[\delta
(s^{\prime }+t_{1}+u_{1}) +\sum_{n=1}^{\infty }\left( \frac{\alpha
_{s}C_{A}}{\pi}\right)^{n}K^{(n)}( s^{\prime },t_{1},u_{1})\right].
\end{equation}

The functions $K^{(n)}( s^{\prime },t_{1},u_{1}) $ in Eq.~(\ref{16}) originate from the collinear and soft limits. Since the azimuthal angle $\varphi $ is the same for both $\gamma ^{*}g$ and $Q\bar{Q}$ center-of-mass systems in these limits, the functions 
$K^{(n)}( s^{\prime },t_{1},u_{1}) $ are also the same for all $\hat{\sigma}_{k}$, $k=T,L,A,I$. 
At NLO, the soft-gluon corrections to NLL accuracy in the $\overline{\text{MS}}$ scheme read  \cite{Laenen-Moch}
\begin{eqnarray}
K^{(1)}( s^{\prime },t_{1},u_{1}) &=& 2\left[ \frac{\ln \left( s_{4}/m^{2}\right)
}{s_{4}}\right]_{+}-\left[ \frac{1}{s_{4}}\right]_{+}\left[ 1+\ln
\frac{u_{1}}{t_{1}} -\left( 1-\frac{2C_{F}}{ C_{A}}\right) \left(
1+\text{Re}L_{\beta }\right) +\ln \frac{\mu ^{2}}{m^{2}} \right]   \nonumber \\
&&{}+\delta ( s_{4}) \ln \frac{-u_{1}}{m^{2}} \ln \frac{\mu ^{2}}{m^{2}}. \label{17}
\end{eqnarray}
In Eq.~(\ref{17}), $C_{A}=N_{c}$, $ C_{F}=(N_{c}^{2}-1)/(2N_{c})$, $N_{c}$ is the number of quark colors, and $ L_{\beta }=(1-2m^{2}/s)\{\ln[(1-\beta_{z})/(1+\beta_{z})]+$i$\pi\}$ with $\beta_{z}=\sqrt{1-4m^{2}/s}$. The single-particle inclusive ``plus'' distributions are defined by
\begin{equation}  \label{18}
\left[\frac{\ln^{l}\left( s_{4}/m^{2}\right) }{s_{4}}\right]_{+}=\lim_{\epsilon
\rightarrow 0}\left[\frac{\ln^{l}\left(s_{4}/m^{2}\right) }{s_{4}}\theta (
s_{4}-\epsilon)+\frac{1}{l+1}\ln ^{l+1}\frac{\epsilon }{m^{2}} \delta (
s_{4})\right].
\end{equation}
For any sufficiently regular test function $h(s_{4})$, Eq.~(\ref{18}) implies that
\begin{equation} \label{19}
\int_{0}^{s_{4}^{\max }}\text{d}s_{4}\,h(s_{4})\left[ \frac{\ln ^{l}\left(
s_{4}/m^{2}\right) }{s_{4}}\right]_{+}=\int_{0}^{s_{4}^{\max
}}\text{d}s_{4}\left[ h(s_{4})-h(0)\right] \frac{\ln ^{l}\left( s_{4}/m^{2}\right)
}{s_{4}}+\frac{1}{l+1}h(0)\ln ^{l+1}\frac{s_{4}^{\max }}{m^{2}}.
\end{equation}

Standard NLL soft-gluon approximation allows us to determine unambiguously only the singular $s_{4}$ behavior of the cross sections defined by Eq.~(\ref{18}). To fix the $s_{4}$ dependence of the Born-level distributions $B_{k}^{\text{{\rm Born}}} (s^{\prime},t_{1},u_{1})\bigr|_{u_{1}=s_{4}-s^{\prime}-t_{1}}$ in Eq.~(\ref{16}), we use the method  proposed in \cite{we7} and based on comparison of the soft-gluon predictions with the exact NLO results. According to \cite{we7},
\begin{equation}\label{20}
B_{k}^{\text{{\rm Born}}}( s^{\prime },t_{1},u_{1}) \equiv s^{\prime 2}\frac{\text{d}\hat{\sigma}^{(0)}_{k,g}}{\text{d}t_{1}}(x_{4}s^{\prime},x_{4}t_{1}), \qquad \qquad x_{4}=-\frac{u_{1}}{s^{\prime}+t_{1}}=1-\frac{s_{4}}{s^{\prime}+t_{1}},
\end{equation}
where the LO GF differential distributions 
$s^{\prime 2}\text{d}\hat{\sigma}^{(0)}_{k,g}(s^{\prime},t_{1})/\text{d}t_{1}$ are
\begin{eqnarray}
s^{\prime 2}\frac{\text{d}\hat{\sigma}^{(0)}_{T,g}}{\text{d}t_{1}}(s^{\prime},t_{1})&=&\pi e_{Q}^{2}\alpha_{\mathrm{em}}\alpha_{s}\left.\left\{\frac{t_{1}}{u_{1}}+\frac{u_{1}}{t_{1} }+4\left( \frac{s}{s^{\prime}}-\frac{m^{2}s^{\prime }}{t_{1}u_{1}}\right) \left[ \frac{s^{\prime}(m^{2}-Q^{2}/2)}{t_{1}u_{1}}+\frac{Q^{2}}{s^{\prime}}\right] \right\}\right|_{u_{1}=-s^{\prime}-t_{1}} \nonumber \\
s^{\prime 2}\frac{\text{d}\hat{\sigma}^{(0)}_{L,g}}{\text{d}t_{1}}(s^{\prime},t_{1})&=&\pi e_{Q}^{2}\alpha_{\mathrm{em}}\alpha_{s}\left.\frac{8Q^{2}}{s^{\prime }}\left( \frac{s}{s^{\prime }}-\frac{m^{2}s^{\prime }}{t_{1}u_{1}}\right)\right|_{u_{1}=-s^{\prime}-t_{1}} \nonumber  \\
s^{\prime 2}\frac{\text{d}\hat{\sigma}^{(0)}_{A,g}}{\text{d}t_{1}}(s^{\prime},t_{1})&=&\pi e_{Q}^{2}\alpha_{\mathrm{em}}\alpha_{s}\left.4\left( \frac{s}{s^{\prime }}-\frac{ m^{2}s^{\prime }}{t_{1}u_{1}}\right) \left(\frac{m^{2}s^{\prime }}{t_{1}u_{1}}+\frac{Q^{2}}{s^{\prime }}\right) \right|_{u_{1}=-s^{\prime}-t_{1}} \label{21}  \\
s^{\prime 2}\frac{\text{d}\hat{\sigma}^{(0)}_{I,g}}{\text{d}t_{1}}(s^{\prime},t_{1})&=&\pi e_{Q}^{2}\alpha_{\mathrm{em}}\alpha_{s}\left.4\sqrt{Q^{2}}\left( \frac{t_{1}u_{1}s }{s^{\prime 2}}-m^{2}\right)^{1/2}\frac{t_{1}-u_{1}}{t_{1}u_{1}}\left( 1-\frac{2Q^{2}}{s^{\prime }}-\frac{2m^{2}s^{\prime}}{t_{1}u_{1}}\right) \right|_{u_{1}=-s^{\prime}-t_{1}}    \nonumber 
\end{eqnarray}

Comparison with the exact NLO results given by Eqs.~(4.7) and (4.8) in Ref.~\cite{LRSN} indicates that the usage of the distributions $B_{k}^{\text{{\rm Born}}}(s^{\prime},t_{1},u_{1})$ defined by Eqs.~(\ref{20}) and (\ref{21}) in present paper provides an accurate account of the logarithmic contributions originating from collinear gluon emission. 
Numerical analysis shows that Eqs.~(\ref{20}) and (\ref{21}) render it possible to describe with good accuracy the exact NLO predictions for the functions $\hat{\sigma}^{(1)}_{T}(s^{\prime})$ and $\hat{\sigma}^{(1)}_{L}(s^{\prime})$  near the threshold at relatively low virtualities $Q^{2}\sim m^{2}$  \cite{we7}.\footnote{Note that soft-gluon approximation is unreliable for
high $Q^{2} \gg m^{2}$.}
\begin{figure}[t]
\begin{center}
\begin{tabular}{cc}
\mbox{\epsfig{file=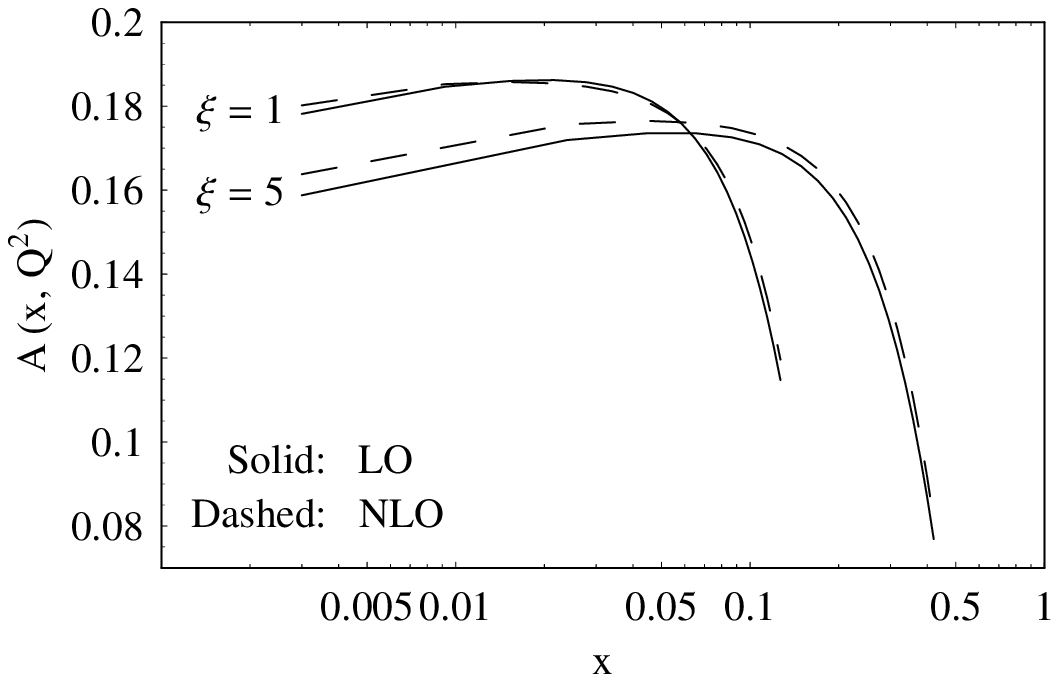,width=210pt}}
& \mbox{\epsfig{file=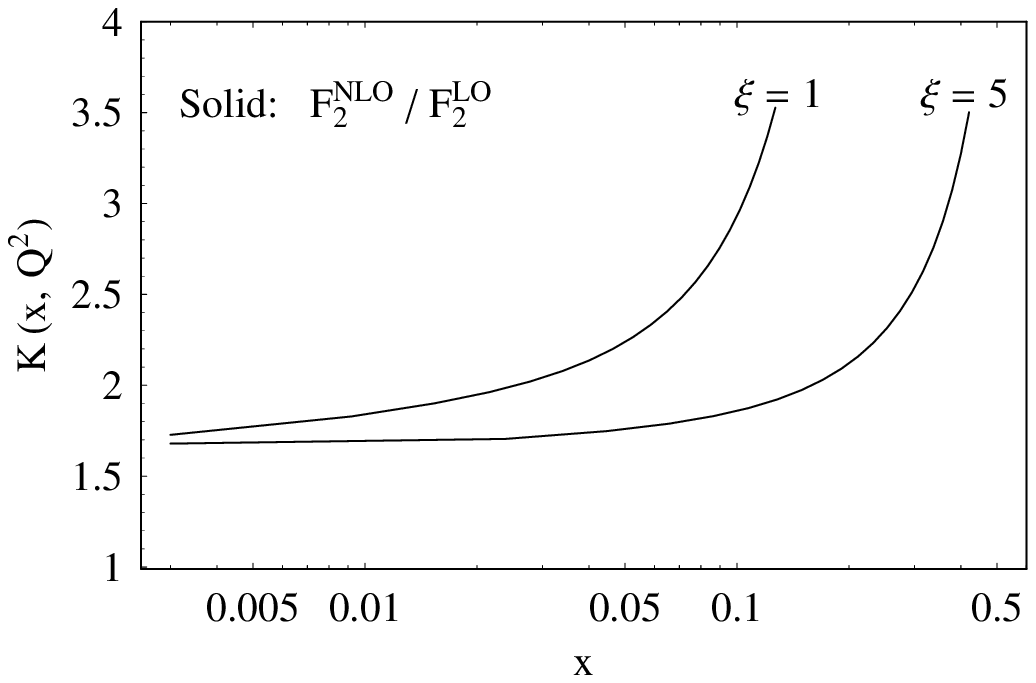,width=210pt}}\\
\end{tabular}
\caption{\label{Fg.4}\small \emph{Left panel:} LO (solid lines) and NLO (dashed lines)  soft-gluon predictions for the $x$ dependence of the azimuthal $\cos(2\varphi)$ asymmetry,
$A(x,Q^{2})=2xF_{A}/F_{2}$, in charm leptoproduction at $\xi=1$ and 5.
\emph{Right panel:} $x$ dependence of the $K$ factor,  $K(x,Q^2)=F_2^{\mathrm{NLO}}/F_2^{\mathrm{LO}}$, at the same values of $\xi$.}
\end{center}
\end{figure}

Our results for the $x$ distribution of the azimuthal $\cos(2\varphi)$ asymmetry, $A(x,Q^{2})=2xF_{A}/F_{2}$, in charm leptoproduction at fixed values of $\xi$ are presented in the left panel of Fig.~\ref{Fg.4}. For comparison, the $K$ factor,  
$K(x,Q^2)=F_2^{\mathrm{NLO}}/F_2^{\mathrm{LO}}$, for the structure function $F_2$ at the same values of $\xi$ is shown in the right panel of Fig.~\ref{Fg.4}. One
can see that the sizable soft-gluon corrections to the production cross sections affect
the Born predictions for $A(x,Q^2)$ at NLO very little, by a few percent only.

\section{\label{analytic} Analytic LO results for $R(x,Q^2)$ and $A(x,Q^2)$ at low $x$}

Since the ratios $R(x,Q^2)$ and $A(x,Q^2)$ are perturbatively stable, it makes sense to provide
the LO hadron-level predictions for these quantities in analytic form. In this Section, we derive compact low-$x$ approximation formulae for the azimuthal $\cos(2\varphi)$ asymmetry and the quantity $R_2(x,Q^2)$ closely related to the Callan-Gross ratio $R(x,Q^2)$,
\begin{equation}  \label{22}
R_{2}(x,Q^{2})=2x\frac{F_{L}}{F_{2}}(x,Q^2)=\frac{R(x,Q^{2})}{1+R(x,Q^{2})}.
\end{equation}
We will see below that our obtained results may be useful in the extraction of the structure functions  $F_k$  ($k=2,L,A,I$) from experimentally measurable reduced cross sections. 

To obtain the hadron-level predictions, we convolute the LO partonic cross sections given by Eqs.~(\ref{8}) with the low-$x$ asymptotics of the gluon PDF:
\begin{equation} \label{23}
g(x,Q^2)\stackrel{x\to 0}{\longrightarrow}\frac{1}{x^{1+\delta}}.
\end{equation}

The value of $\delta$ in Eq.~(\ref{23}) is a matter of discussion. The simplest choice,
$\delta =0$, leads to a non-singular behavior of the structure functions for $x\to 0$.
Another extreme value, $\delta =1/2$, historically originates from the BFKL resummation 
of the leading powers of $\ln(1/x)$ \cite{BFKL1,BFKL2,BFKL3}. In reality, $\delta$ is a function 
of $Q^2$ (for an experimental review, see Refs.~\cite{A-Vogt,PDF-LHC}). Theoretically, the $Q^2$ 
dependence of $\delta$ is calculated using the DGLAP evolution equations \cite{DGLAP1,DGLAP2,DGLAP3}.

First, we derive an analytic low-$x$ formula for the ratio $R_{2}^{(\delta )}(Q^2)\equiv R^{(\delta)}_2(x\to 0,Q^2)$ with arbitrary values of $\delta$ in terms of the Gauss hypergeometric function.
Our result has the following form:
\begin{equation} \label{24}
R_{2}^{(\delta )}(Q^2)=4\frac{\frac{2+\delta }{3+\delta }\Phi
\left( 1+\delta ,\frac{1}{1+4\lambda }\right) -\left( 1+4\lambda \right)
\Phi \left( 2+\delta ,\frac{1}{1+4\lambda }\right) }{\left[ 1+\frac{%
\delta \left( 1-\delta ^{2}\right) }{\left( 2+\delta \right) \left( 3+\delta \right)
}\right] \Phi \left( \delta ,\frac{1}{1+4\lambda }\right) -\left( 1+4\lambda \right)
\left( 4-\delta -\frac{10}{3+\delta }\right) \Phi \left( 1+\delta ,\frac{1}{1+4\lambda
}\right) },
\end{equation}
where $\lambda$ is defined in Eq.~(\ref{9}) and the function $\Phi (r,z)$ is 
\begin{equation} \label{25}
\Phi \left( r,z\right) =\frac{z^{1+r}}{1+r}\,\frac{\Gamma \left( 1/2\right) \Gamma \left(
1+r\right) }{\Gamma \left( 3/2+r\right) }\,{}_{2}F_{1}\left(
\frac{1}{2},1+r;\frac{3}{2}+r;z\right) .
\end{equation}
The hypergeometric function ${}_{2}F_{1}(a,b;c;z)$ has the following series expansion:
\begin{equation} \label{26}
{}_{2}F_{1}\left( a,b;c;z\right)=\frac{\Gamma \left( c\right) }{\Gamma \left( a\right)
\Gamma \left( b\right)}\sum\limits_{n=0}^{\infty }\frac{\Gamma
\left( a+n\right) \Gamma \left( b+n\right) }{\Gamma \left( c+n\right) }\frac{%
z^{n}}{n!}. 
\end{equation}

\begin{figure}[t]
\begin{center}
\begin{tabular}{cc}
\mbox{\epsfig{file=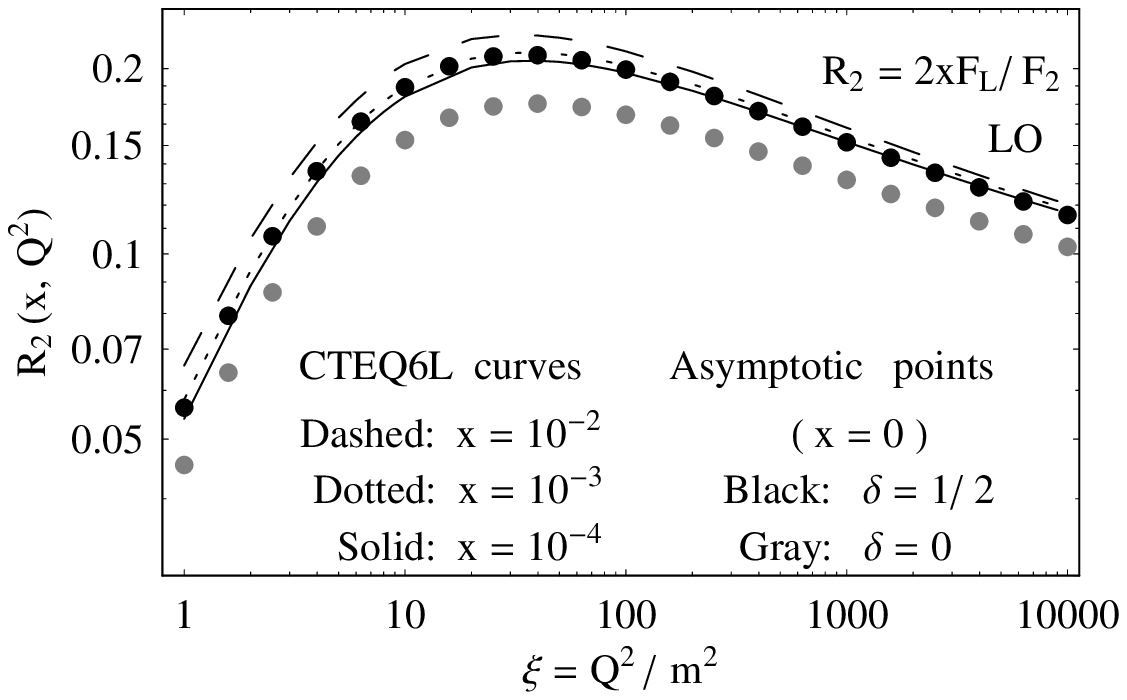,width=210pt}}
& \mbox{\epsfig{file=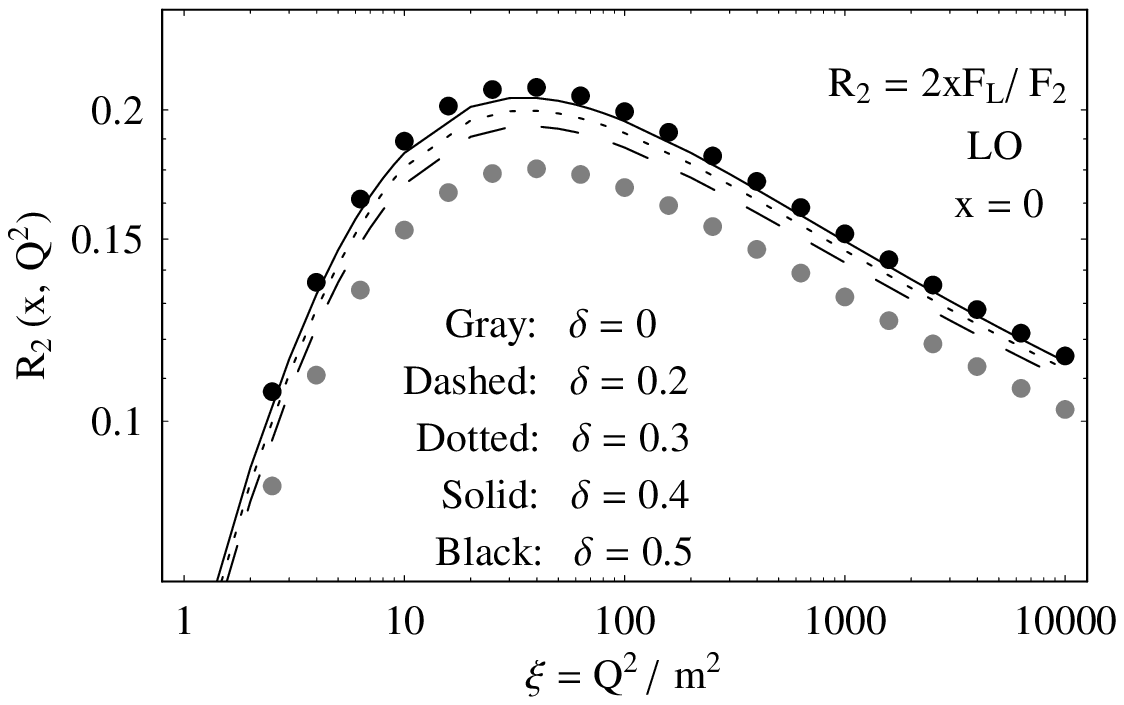,width=210pt}}\\
\end{tabular}
 \caption{\label{Fg.5}\small LO low-$x$ predictions for the ratio $R_2(x,Q^2)=2xF_L/F_2$
 in charm leptoproduction. \emph{Left panel:} Asymptotic ratios $R^{(0)}_2(Q^2)$
 (gray points) and $R^{(1/2)}_2(Q^2)$ (black points), as well as CTEQ6L
 predictions for $R_2(x,Q^2)$ at $x=10^{-2}$, $10^{-3}$ and $10^{-4}$.
 \emph{Right panel:}
 Asymptotic ratio $R^{(\delta )}_2(Q^2)$ at
 $\delta=0$, 0.2, 0.3, 0.4 and 0.5.}
\end{center}
\end{figure}

In Fig.~\ref{Fg.5}, we investigate the obtained result (\ref{24}) for  $R_{2}^{(\delta )}(Q^2)$. The left panel of Fig.~\ref{Fg.5} shows the ratio $R_{2}^{(\delta )}(Q^2)$ as functions of $\xi$ for two  extreme cases, $\delta =0$ and $1/2$. One can see that the difference between these
quantities varies slowly from $20\%$ at low $Q^2$ to $10\%$ at high $Q^2$. For
comparison, also the LO results for $R_2(x,Q^2)$ calculated at several values of $x$
using the CTEQ6L gluon PDF \cite{CTEQ6} are shown. We observe that, for $x\to 0$, the
CTEQ6L predictions converge to the function $R^{(1/2)}_2(Q^2)$ practically in the entire
region of $Q^2$. We have verified that the similar situation takes also place for other recent 
CTEQ PDF versions \cite{CTEQ6,CT10}.
In the right panel of Fig.~\ref{Fg.5}, the $\delta$ dependence of the asymptotic ratio
$R^{(\delta)}_2(Q^2)$ is investigated. One can see that the ratio $R^{(\delta)}_2(Q^2)$
rapidly converges to the function $R^{(1/2)}_2(Q^2)$ for $\delta > 0.2$. In particular,
the relative difference between $R^{(0.5)}_2(Q^2)$ and $R^{(0.3)}_2(Q^2)$ varies slowly
from $6\%$ at low $Q^2$ to $2\%$ at high $Q^2$. 

As mentioned above, the $Q^2$ dependence of the parameter $\delta$ is determined with the 
help of the DGLAP evolution. However, our analysis shows that hadron-level predictions for 
$R^{(\delta)}_2(x\to 0,Q^2)$ depend weakly on $\delta$ practically in the entire region of $Q^2$ 
for $0.2< \delta < 0.9$. For this reason, it makes sense to consider the ratio $R^{(\delta)}_2(Q^2)$ in particular case of $\delta = 1/2$. The result is:
\begin{equation} \label{27}
R^{(1/2)}_2(Q^2)=\frac{8}{1 + 4\lambda}\,\frac{ \left[ 3 + 4\lambda \left( 13
+ 32\lambda \right) \right] E(1/(1 + 4\lambda)) - 4\lambda \left( 9 +
32\lambda  \right) K(1/(1 + 4\lambda)) }{ \left( -37 + 72\lambda
\right)E(1/(1 + 4\lambda)) + 2\left( 23 - 36\lambda
\right)K(1/(1 + 4\lambda)) },
\end{equation}
where the functions $K(y)$ and $E(y)$ are the complete elliptic integrals of the first and
second kinds defined as
\begin{equation} \label{28}
K(y)=\int_0^1\frac{{\text d}t}{\sqrt{(1-t^2)(1-yt^2)}},
\qquad
E(y)=\int_0^1{\text d}t \sqrt{\frac{1-yt^2}{1-t^2}}.
\end{equation}

One can see from Fig.~\ref{Fg.5} that our simple formula (\ref{27}) with $\delta =1/2$ 
(i.e., without any evolution) describes with good accuracy the low-$x$ CTEQ results for 
$R_2(x,Q^2)$. We conclude that the hadron-level predictions for $R_2(x\to 0,Q^2)$ are stable not only under the NLO corrections to the partonic cross sections, but also under the DGLAP evolution of the gluon PDF.

Then we calculate and investigate the LO hadron-level predictions for the azimuthal $\cos(2\varphi)$ asymmetry in the limit of $x\to 0$. Our result for the quantity $A^{(\delta )}(Q^2)\equiv A^{(\delta)}(x\to 0,Q^2)$  has the following form:
\begin{equation} \label{29}
A^{(\delta )}(Q^2)=2\frac{\frac{2+\delta +2\lambda}{3+\delta }\Phi
\left( 1+\delta ,\frac{1}{1+4\lambda }\right) -\left( 1+4\lambda \right)
\Phi \left( 2+\delta ,\frac{1}{1+4\lambda }\right) }{\left[ 1+\frac{%
\delta \left( 1-\delta ^{2}\right) }{\left( 2+\delta \right) \left( 3+\delta \right)
}\right] \Phi \left( \delta ,\frac{1}{1+4\lambda }\right) -\left( 1+4\lambda \right)
\left( 4-\delta -\frac{10}{3+\delta }\right) \Phi \left( 1+\delta ,\frac{1}{1+4\lambda
}\right) }.
\end{equation}

\begin{figure}[t]
\begin{center}
\begin{tabular}{cc}
\mbox{\epsfig{file=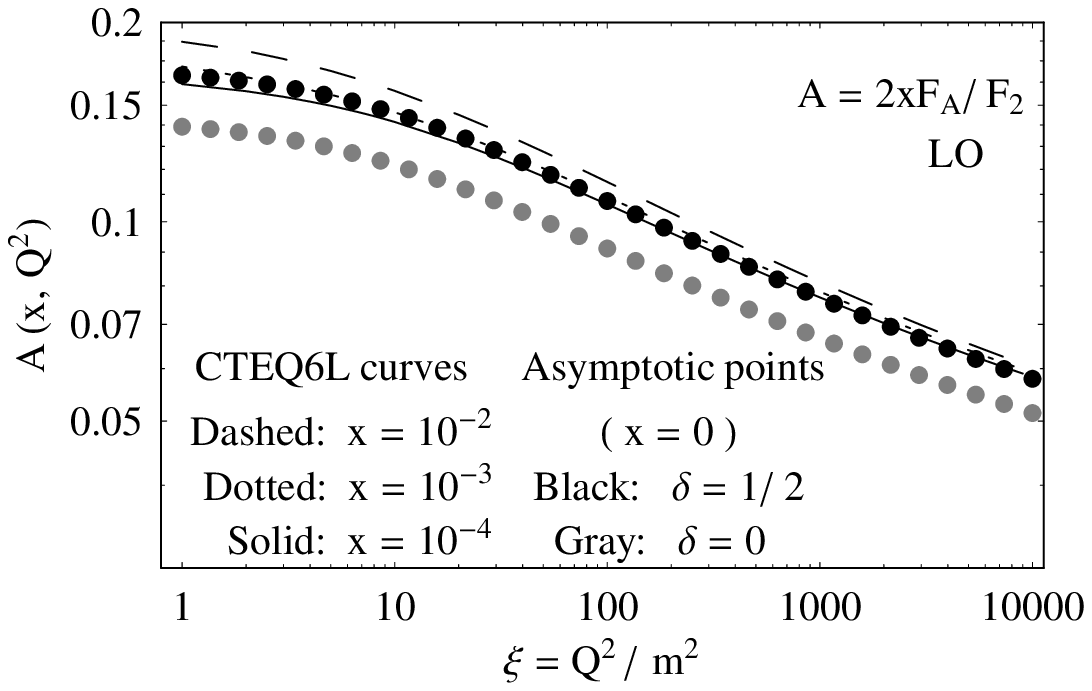,width=210pt}}
& \mbox{\epsfig{file=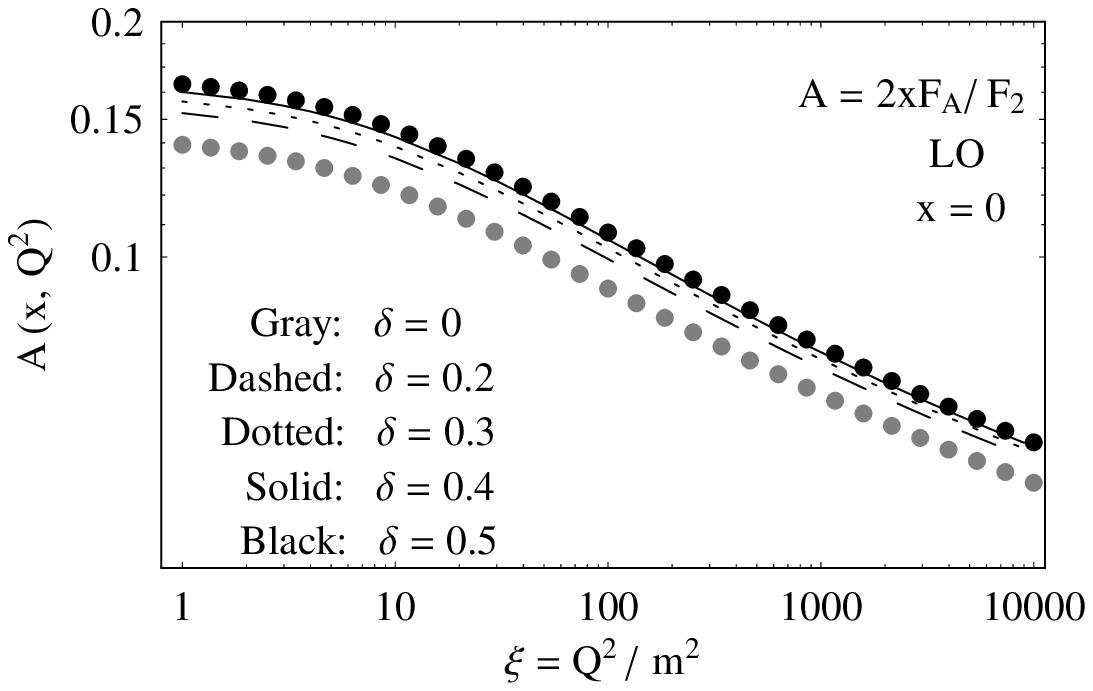,width=210pt}}\\
\end{tabular}
 \caption{\label{Fg.6}\small LO low-$x$ predictions for the ratio $A(x,Q^2)=2xF_A/F_2$
 in charm leptoproduction. \emph{Left panel:} Asymptotic ratios $A^{(0)}(Q^2)$
 (gray points) and $A^{(1/2)}(Q^2)$ (black points), as well as CTEQ6L
 predictions for $A(x,Q^2)$ at $x=10^{-2}$, $10^{-3}$ and $10^{-4}$.
 \emph{Right panel:}
 Asymptotic ratio $A^{(\delta )}(Q^2)$ at
 $\delta=0$, 0.2, 0.3, 0.4 and 0.5.}
\end{center}
\end{figure}

Our analysis presented in Fig.~\ref{Fg.6} shows that the quantity $A^{(\delta )}(Q^2)$ defined by Eq.~(\ref{29}) has the properties very similar to the ones demonstrated by the ratio $R^{(\delta)}_2(Q^2)$. In particular, one can see from Fig.~\ref{Fg.6} that the hadron-level predictions for $A^{(\delta )}(Q^2)$ depend weakly on $\delta$ practically in the entire region of $Q^2$ for $\delta > 0.2$. So, the azimuthal $\cos(2\varphi)$ asymmetry $A(x\to 0,Q^2)$ is also stable under the DGLAP evolution of the gluon PDF.

Let us now discuss how the obtained analytic results may be used in the extraction of the structure functions  $F_k$  ($k=2,L,A,I$) from experimentally measurable quantities. Usually,  it is the so-called "reduced cross section", $\tilde{\sigma}(x,Q^{2})$, that can directly be measured in DIS experiments:
\begin{eqnarray}
\tilde{\sigma}(x,Q^{2})=\frac{1}{1+(1-y)^2}\frac{xQ^4}{2\pi
\alpha^{2}_{\mathrm{em}}}\frac{\text{d}^{2}\sigma_{lN}}{\text{d}x\text{d}Q^{2}}&=&F_{2}(
x,Q^{2})-\frac{2xy^{2}}{1+(1-y)^2}F_{L}(x,Q^{2}) \label{30} \\
&=&F_{2}(x,Q^{2})\left[1-\frac{y^{2}}{1+(1-y)^2}R_{2}(x,Q^{2}) \right]. \label{31}
\end{eqnarray} 
In earlier HERA analyses of charm and bottom electroproduction, the corresponding longitudinal structure functions were taken to be zero for simplicity. In this case, $\tilde{\sigma}(x,Q^{2})=F_{2}(x,Q^{2})$. In recent papers \cite{H1HERA1,H1HERA2}, the structure function $F_{2}(x,Q^2)$ is evaluated from the reduced cross section (\ref{30}) where the longitudinal structure function $F_{L}(x,Q^2)$ is estimated from the NLO QCD expectations. Instead of this rather cumbersome procedure, we propose to use the expression (\ref{31}) with the quantity $R_{2}(x,Q^2)$ defined by the  analytic LO expressions (\ref{24}) or (\ref{27}). This simplifies the extraction of $F_{2}(x,Q^2)$ from measurements of $\tilde{\sigma}(x,Q^{2})$ but does not affect the accuracy of the result in practice because of perturbative stability of the ratio $R_{2}(x,Q^2)$.

In Ref.~\cite{we7}, we used the analytic expressions (\ref{24}) and (\ref{27}) for the
extraction of the structure functions $F_2^c(x,Q^2)$ and $F_2^b(x,Q^2)$ from the HERA
measurements of the reduced cross sections $\tilde{\sigma}^c(x,Q^2)$ and
$\tilde{\sigma}^b(x,Q^2)$, respectively. It was demonstrated that our LO formula (\ref{27}) for $R_{2}(x,Q^2)$ with $\delta=1/2$ usefully reproduces the results for 
$F_2^c(x,Q^2)$ and $F_2^b(x,Q^2)$ obtained by the H1 Collaboration \cite{H1HERA1,H1HERA2} 
with the help of the NLO evaluation of $F_{L}(x,Q^2)$. In particular, the results of our analysis of the HERA data on the charm electroproduction are collected in Table~\ref{tab1}. In our calculations, the value $m_c=1.3$~GeV for the charm quark mass is used. The LO predictions, $F_2(\mathrm{LO})$, for the case of $\delta=0.5$ are presented and compared with the NLO values, $F_2(\mathrm{NLO})$, obtained in the H1 analysis \cite{H1HERA1,H1HERA2}. One can see that our LO predictions agree with the NLO results with an accuracy better than 1\%.

\begin{table}[h]
\caption{\label{tab1} Values of $F_2^c(x,Q^2)$ extracted from the HERA measurements of
$\tilde{\sigma}^c(x,Q^{2})$ at low \cite{H1HERA2} and high \cite{H1HERA1} $Q^2$ (in
GeV$^2$) for various values of $x$ (in units of $10^{-3}$). The NLO H1 results \cite{H1HERA1,H1HERA2} are compared with the LO predictions corresponding to
the case of $\delta =0.5$.}
\begin{center}
\begin{tabular}{||ccc||cc||cc||}
\hline
 $\quad Q^2 \quad$ & $x$ & ~$\quad y\quad$~ & ~$\quad \tilde{\sigma}^c \quad$~ &
  ~Error~ &
$\qquad F_2^c$(NLO)$\qquad$ & $ F_2^c$(LO)  \\
 (GeV$^2$) & $(\times 10^{-3})$  &  &  & (\%) & H1 & $\delta=0.5$ \\
\hline
  \hline
 12 & 0.197 & 0.600 & 0.412 & 18 & $0.435\pm0.078$ & $0.435\pm0.078$  \\
 12 & 0.800 & 0.148 & 0.185 & 13 & $0.186\pm0.024$ & $0.185\pm0.024$  \\
 25 & 0.500 & 0.492 & 0.318 & 13 & $0.331\pm0.043$ & $0.331\pm0.043$  \\
 25 & 2.000 & 0.123 & 0.212 & 10 & $0.212\pm0.021$ & $0.212\pm0.021$  \\
 60 & 2.000 & 0.295 & 0.364 & 10 & $0.369\pm0.040$ & $0.369\pm0.040$  \\
 60 & 5.000 & 0.118 & 0.200 & 12 & $0.201\pm0.024$ & $0.200\pm0.024$  \\
200 & 0.500 & 0.394 & 0.197 & 23 & $0.202\pm0.046$ & $0.202\pm0.046$  \\
200 & 1.300 & 0.151 & 0.130 & 24 & $0.131\pm0.032$ & $0.130\pm0.031$  \\
650 & 1.300 & 0.492 & 0.206 & 27 & $0.213\pm0.057$ & $0.213\pm0.057$  \\
650 & 3.200 & 0.200 & 0.091 & 31 & $0.092\pm0.028$ & $0.091\pm0.028$  \\
\hline
\end{tabular}
\end{center}
\end{table}

The structure functions $F_A$ and $F_I$ can be extracted from the $\varphi$-dependent DIS cross section,
\begin{align}
\frac{\mathrm{d}^{3}\sigma_{lN}}{\mathrm{d}x\mathrm{d}Q^{2}\mathrm{d}\varphi}
=\frac{2\alpha^{2}_{em}}{Q^4}\frac{y^2}{1-\varepsilon}\Bigl[\frac{1}{2x}& F_{2}( x,Q^{2})- (1-\varepsilon) F_{L}(x,Q^{2}) \Bigr. \label{32} \\
+\Bigl. \varepsilon & F_{A}( x,Q^{2})\cos 2\varphi+2\sqrt{\varepsilon(1+\varepsilon)} F_{I}( x,Q^{2})\cos \varphi\Bigr], \nonumber 
\end{align}
where $\varepsilon=\frac{2(1-y)}{1+(1-y)^2}$. 
For this purpose, one should measure the first moments of the $\cos(\varphi)$ and $\cos(2\varphi)$ distributions defined as
\begin{equation}  \label{33}
\langle \cos n\varphi \rangle (x,Q^{2})= \frac{\int_{0}^{2\pi }\text{d}\varphi \cos
n\varphi {\text{d}^{3}\sigma _{lN} \over \text{d}x\text{d}Q^{2}\text{d}\varphi } (x,Q^{2},\varphi ) }{\int_{0}^{2\pi }\text{d}\varphi {\text{d}^{3}\sigma_{lN} \over \text{d}x\text{d}Q^{2}\text{d}\varphi } (x,Q^{2},\varphi ) }. 
\end{equation}
Using Eq.~(\ref{32}), we obtain:
\begin{align} 
\langle \cos 2\varphi \rangle(x,Q^{2})&=\frac{1}{2}\frac{\varepsilon A(x,Q^{2})}{1-(1-\varepsilon)R_2(x,Q^{2})},& A(x,Q^{2})&=2x\frac{F_{A}}{F_{2}}(x,Q^{2}),\label{34}\\
\langle \cos \varphi \rangle(x,Q^{2})&=\frac{\sqrt{\varepsilon (1+\varepsilon)} A_I(x,Q^{2})}{1-(1-\varepsilon)R_2(x,Q^{2})}, & A_I(x,Q^{2})&=2x\frac{F_{I}}{F_{2}}(x,Q^{2}).\label{35}
\end{align}

One can see from Eqs.~(\ref{34}) and (\ref{35}) that, using the perturbatively stable predictions (\ref{24}) for $R_2(x,Q^{2})$, we will be able to determine the structure functions $F_A(x,Q^{2})$ and $F_I(x,Q^{2})$ from future data on the moments $\langle\cos 2\varphi\rangle$ and $\langle\cos \varphi\rangle$. On the other hand, according to Eq.~(\ref{34}), the analytic results (\ref{24}) and (\ref{29}) for the quantities $R_2(x,Q^{2})$ and $A(x,Q^{2})$ provide us with the perturbatively stable predictions for $\langle\cos 2\varphi\rangle$ which may be directly tested in experiment.

So, our obtained analytic and perturbatively stable predictions for the ratios $R_2(x,Q^{2})$ and $A(x,Q^{2})$ will simplify both the extraction of structure functions from the measurable  $\varphi$-dependent cross section (\ref{32}) and the test of self-consistency of the extraction procedure.   

\section{Conclusion}

We conclude by summarizing our main observations. In the present paper, we studied
the radiative corrections to the Callan-Gross ratio, $R(x,Q^{2})$, and azimuthal $\cos(2\varphi)$ asymmetry, $A(x,Q^{2})$, in heavy-quark leptoproduction. 
It turned out that large 
(especially, at non-small $x$) radiative corrections to the structure functions
cancel each other in their ratios $R(x,Q^2)=F_L/F_T$ and $A(x,Q^{2})=2xF_A/F_A$ with good accuracy. As a result, the NLO contributions to the ratios $R(x,Q^{2})$ and $A(x,Q^{2})$ are less than $10\%$ in a wide region of the variables $x$ and $Q^2$. Our analysis shows that, sufficiently above the production threshold, the pQCD predictions for $R(x,Q^2)$ and $A(x,Q^{2})$ are insensitive (to within ten percent) to standard uncertainties in the QCD input parameters and to the DGLAP evolution of PDFs. We conclude that, unlike the production cross sections, the Callan-Gross ratio and $\cos(2\varphi)$ asymmetry in heavy-quark leptoproduction are quantitatively well defined in pQCD. Measurements of the quantities $R(x,Q^2)$ and $A(x,Q^{2})$ in charm and bottom leptoproduction would provide a good test of the conventional parton model based on pQCD.

As to the experimental aspects, we propose to exploit the observed perturbative
stability of the Callan-Gross ratio and azimuthal asymmetry in the extraction of the structure functions from the experimentally measurable reduced cross sections. For this purpose, we provided compact LO hadron-level formulae for the ratios $R_{2}(x,Q^2)=2xF_L/F_2=R/(1+R)$ and $A(x,Q^2)=2xF_A/F_2$ in the limit $x\to 0$. We demonstrated that these analytic expressions usefully reproduce the results for $F_2^c(x,Q^2)$ and $F_2^b(x,Q^2)$ obtained by the H1 Collaboration \cite{H1HERA1,H1HERA2} with the help of the more cumbersome NLO evaluation of $F_{L}(x,Q^2)$. Our obtained predictions will also be useful in extraction of the azimuthal asymmetries from the incoming COMPASS results as well as from future data on heavy-quark leptoproduction at the proposed EIC \cite{EIC} and LHeC \cite{LHeC} colliders at BNL/JLab and CERN, correspondingly.

\acknowledgments{
The author is thankful to Serge Bondarenko for invitation to XXI International Baldin Seminar on High Energy Physics Problems and help. We thank S.~I.~Alekhin and J.~Bl\"umlein for providing us with fast code \cite{Blumlein} for numerical calculations of the NLO partonic cross sections. The author is also grateful to S.~J.~Brodsky, A.~V.~Efremov, A.~V.~Kotikov, A.~B.~Kniehl, E.~Leader and C.~Weiss for useful discussions. This work is supported in part by the State Committee of Science of RA, grant 11-1C015.}

\end{document}